\documentclass{article}

\usepackage{arxiv}
\usepackage{color}
\usepackage[utf8]{inputenc} 
\usepackage[T1]{fontenc}    
\usepackage[hidelinks]{hyperref}       
\usepackage{url}            
\usepackage{booktabs}       
\usepackage{amsfonts}       
\usepackage{nicefrac}       
\usepackage{microtype}      
\usepackage{lipsum}		
\usepackage{float}
\usepackage{graphicx}
\usepackage[ruled,vlined]{algorithm2e}

\usepackage[final]{changes}
\usepackage{amsmath}
\usepackage{amsfonts}

\usepackage{multirow}
\usepackage{adjustbox}
\usepackage{booktabs}
\newcommand{\ra}[1]{\renewcommand{\arraystretch}{#1}}
\newtheorem{define}{Definition}[section] 

\title{U-vectors: Generating clusterable speaker embedding from unlabeled data}

\date{} 					

\author{ 				
	\href{https://orcid.org/0000-0001-5738-1631}{\includegraphics[scale=0.06]{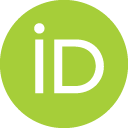}\hspace{1mm}M. F. Mridha} \\
	Department of Computer Science \& Engineering\\
	Bangladesh University of Business \& Technology\\ 
	Dhaka, Bangladesh \\
	\texttt{firoz@bubt.edu.bd} \\
	\And
	\href{https://orcid.org/0000-0001-7375-9040}{\includegraphics[scale=0.06]{orcid.png}\hspace{1mm}Abu Quwsar Ohi} \\
	Department of Computer Science \& Engineering\\
	Bangladesh University of Business \& Technology\\ 
	Dhaka, Bangladesh \\
	\texttt{quwsarohi@bubt.edu.bd} \\
	\And
	Muhammad Mostafa Monowar \\
	Department of Information Technology\\
	Faculty of Computing \& Information Technology\\
	King Abdulaziz University \\
	Jeddah-21589, Kingdom of Saudi Arabia \\
	\texttt{mmonowar@kau.edu.sa} \\
	\And
	Md. Abdul Hamid \\
	Department of Information Technology\\
	Faculty of Computing \& Information Technology\\
	King Abdulaziz University \\
	Jeddah-21589, Kingdom of Saudi Arabia \\
	\texttt{mabdulhamid1@kau.edu.sa} \\
	\And
	Md. Rashedul Islam \\
	Department of Computer Science and Engineering\\
	University of Asia Pacific \\
	Dhaka, Bangladesh \\
	\texttt{rashed.cse@gmail.com} \\
	\And
	Yutaka Watanobe \\
	Department of Computer Science and Engineering \\
	University of Aizu \\
	Aizu-Wakamatsu, Japan \\
}

\begin{document}
\maketitle


\begin{abstract}
	Speaker recognition deals with recognizing speakers by their speech. Most speaker recognition systems are built upon two stages, the first stage extracts low dimensional correlation embeddings from speech, and the second performs the classification task. The robustness of a speaker recognition system mainly depends on the extraction process of speech embeddings, which are primarily pre-trained on a large-scale dataset. As the embedding systems are pre-trained, the performance of speaker recognition models greatly depends on domain adaptation policy, which may reduce if trained using inadequate data. This paper introduces a speaker recognition strategy dealing with unlabeled data, which generates clusterable embedding vectors from small fixed-size speech frames. The unsupervised training strategy involves an assumption that a small speech segment should include a single speaker. Depending on such a belief, a pairwise constraint is constructed with noise augmentation policies, used to train AutoEmbedder architecture that generates speaker embeddings. Without relying on domain adaption policy, the process unsupervisely produces clusterable speaker embeddings, termed unsupervised vectors (u-vectors). The evaluation is concluded in two popular speaker recognition datasets for English language, TIMIT, and LibriSpeech. Also, a Bengali dataset is included to illustrate the diversity of the domain shifts for speaker recognition systems. Finally, we conclude that the proposed approach achieves \replaced{satisfactory }{remarkable} performance using pairwise architectures.
\end{abstract}

\keywords{Speaker recognition
	\and Clustering
	\and Twin networks
    \and Deep learning}

\setcounter{section}{0}
\section{Introduction}

Speech is the most engaging and acceptable form of communication among one another. Artificial intelligence (AI) systems are currently continuously targeting and working on various challenges of speech-related topics, including speech recognition, speech segmentation, speaker recognition, speech diarization, and so on. Among the different sub-domains of AI, Deep learning (DL) strategies often perform superior to other techniques. 

\added{The general implementation of DL was mainly conducted on speech recognition systems. DL methods can be trained on speech recognition without the requirement of speech-to-word alignment \cite{graves2006connectionist}. Often such training strategies are defined as an end-to-end method. End-to-end methods can be easily trained from speech transcripts. Therefore, currently, numerous systems are being developed in the speech recognition domain. Speech recognition systems have exciting usages in voice commands, virtual assistants, search engines, speech-to-text processing, etc. Further, numerous automation systems on speech are currently being developed. A speech denoising mechanism removes environmental sounds from speech audio, helping to provide a clean speech \cite{azarang2020review}. Speech synthesis systems artificially enable computers to produce speech sounds \cite{ning2019review}, helping computers to communicate with humans. Speech emotion systems further extract human emotions from speech \cite{fayek2017evaluating}. Thus, computers can apprehend human feelings. Speech segmentation systems can segment a speech into word/phone levels \cite{kamper2017embedded}, helping to identify words and phones from speech. Further, computers can help humans in developing pronunciation \cite{o2018directions}.} Among the various speech-based solutions, speaker recognition has fascinating usage of identifying users by hearing speech. 

Speaker recognition systems are directly involved with biometric identification systems and are suitable for authenticating users remotely by hearing \replaced{voices }{a voice}. In perspective to various biometric systems, such as facial recognition, fingerprint matching, and so on, speaker recognition also has vast usability in numerous domains, including telecom, banking, search optimization, and diarization \cite{ohi2021deep}. Nevertheless, speaker recognition systems suffer difficulties, including speech states, emotional conditions, environmental noise, health conditions, speaking styles, etc. Further, in comparison with supervised speaker recognition approaches, unsupervised and semi-supervised strategies are hardly investigated \cite{kabir2021survey}. Unsupervised and semi-supervised systems resolve the requirement of labeling a vast quantity of speech data.


DL architectures have been extensively investigated for supervised speaker recognition systems. For speaker and speech recognition models, speech spectrograms and mel-frequency cepstral coefficients (MFCC) \cite{tiwari2010mfcc} are used as a preprocessing strategy. For such cases, convolutional neural networks (CNN) are generally implemented \cite{chowdhury2019fusing}. However, current architectures processes raw-audio and extract speaker recognizable features. SincNet \cite{ravanelli2018speaker} improves the feature extraction process from raw audio waves. The architecture fuses sinc functions with CNN that can extract speaker information from low and high cutoff frequencies. AM-MobileNet1D \cite{9207519} further demonstrates that 1D-CNN architectures are sufficient for identifying features from raw audio waveforms. Also, the architecture requires fewer parameters compared to SincNet. Although supervised speaker recognition architectures perform excellently in recognition tasks on a large set of speakers. However, DL strategies require a vast amount of labeled data to operate on speech-related queries.

Generating speech embeddings has been widely observed in the speaker recognition domain \cite{garcia2011analysis,variani2014deep,snyder2018x}. Embedding refers to generating vectors of continuous values. \added{Often, architectures using speaker embeddings are also termed stage-wise architectures \cite{ohi2021deep}.} Currently, unsupervised speaker recognition systems implement domain adaptation policies, mostly fused with embedding vectors \added{\cite{ohi2021deep,bai2021speaker}}. Domain adaptation refers to finding appropriate similarities, where a framework is trained on training data, and tested on a similar yet unseen data. Hence, domain adaptation strategies may perform poorly when the variation of training and the unseen dataset is massive. Further, domain adaptation strategies \replaced{may also }{also may} produce less accuracy if \deleted{, they are} trained on an inadequate dataset with minimal variation \cite{snyder2018x}. Yet, efforts have been made to reduce the interpretation of unseen data over training data, by adding adversarial training \cite{wang2018unsupervised}, improving training policies \cite{garcia2014unsupervised}, covariance adaptation policies \cite{garcia2014improving}, etc. Although these policies improve the robustness, most strategies are still prone to various speech diversions, such as language, speech pattern, age, emotion, and so on.

Currently, in the aspect of DL, embeddings can be generated using triplet \cite{schroff2015facenet} and pairwise loss \cite{ohi2020autoembedder} techniques. In triplet loss architecture, three parallel inputs flow through the network: anchor, negative, and positive. The positive input contains a similar class w.r.t. to the anchor, whereas the negative input contains a different class. \replaced{Comparatively}{However}, in pairwise architecture, a pair of information \replaced{flows either belonging }{is flowed that either belongs} to a single or different class. Triplet architectures have been perceived in speaker feature extraction in supervised practice \cite{zhang2018text}. 

\begin{figure}[H]
	\centering
	\includegraphics[width=0.8\linewidth]{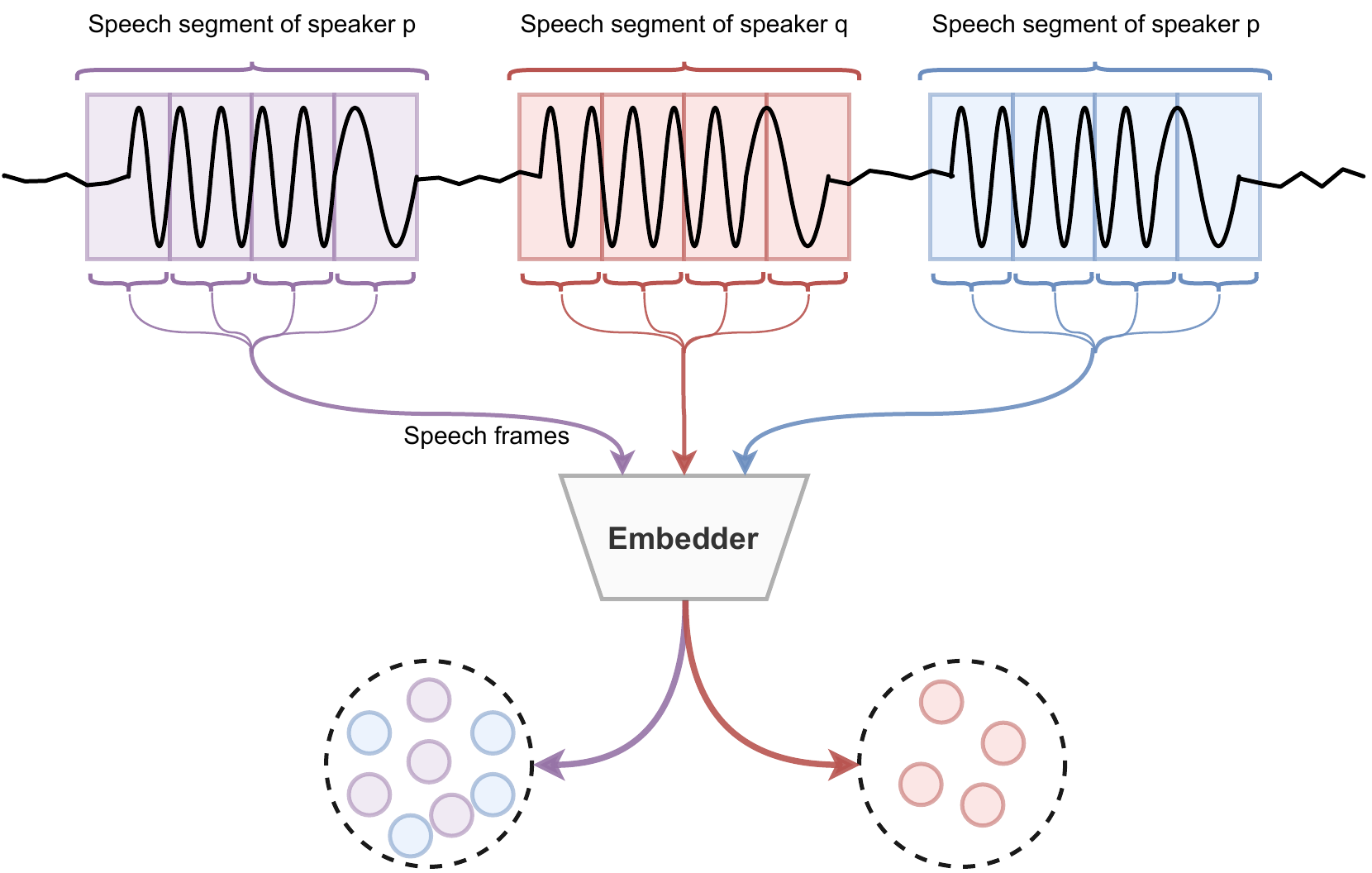}
	\caption{The figure illustrates a set of segmented speech \deleted{is available,} with an unknown number of speakers (in the example, two speakers, p and q). Speech segments are windowed into smaller speech frames, assuming that all frames of a single speech segment belong to a single class. Further, a DL-based embedding system finds speech similarities (inter-segment similarity) and relations from speech segments. The process results in generating clusterable speaker embeddings.}
	\label{fig:speaker_pair}
\end{figure}

This paper introduces an unsupervised strategy of generating speaker embedding directly from the unseen data. Hence, the method does not depend on domain adaptation policies and can adapt diverse features from most speech data. Moreover, we insist on converting DL \replaced{architecture's }{architectures'} training process to both semi-supervised and unsupervised \replaced{manners }{manner}. Yet, to do so, the system requires segmented audio streams (length of 1 second) and needs to guarantee that a segment contains only one person's speech. The audio segment is further windowed into smaller speech frames (0.2 seconds) for training the DL architecture. The audio segments are assigned pseudo labels, which are further reconstructed by DL architecture. Figure \ref{fig:speaker_pair} illustrates the construction of the training procedure.

The overall contributions of the paper are the following:
\begin{itemize}
	\item We introduce \replaced{a }{an} \deleted{unsupervised} strategy of generating speaker-dependent embeddings, named u-vector. The training process is domain-independent and directly learns from the \replaced{unlabeled }{unseen} data.
	\item We use pairwise constraints and non-generative augmentations to train AutoEmbedder architecture. \deleted{using unlabeled data.}
	\item \added{We explore the possibilities of our strategy in both unsupervised and semi-supervised training criteria.}
	\item We evaluate the proposed \replaced{policy }{architecture} with two inter-cluster based strategies: triplet and pairwise architecture\added{s}. 
	\item \added{We finally} conclude that a DL architecture can discriminate speakers from pseudo labels based on feature similarity.
\end{itemize}

We organize the paper as follows, Section \ref{sec:relatedwork} reviews the related works conducted in speaker recognition domain. Section \ref{sec:methodology} clarifies the construction of the training procedure, along with the challenges, and modifications. Section \ref{sec:experiment} illustrates the experimental setup, datasets and the analysis of the \replaced{architecture's}{architectures'} performance. In Section \ref{sec:future_work}, we sketch the proposed method's future initiative\added{s along with usabilities}. Finally, Section \ref{sec:conclusion} concludes the paper.

\section{Related Work}
\label{sec:relatedwork}

Speaker recognition has been a topic of interest over the past decades, and various systems have been proposed to solve the challenge. In the domain of speaker recognition, numerous techniques have been observed since late 2000. Among these, embedding architectures have been widely explored to extract the diversity of speech frames. Embedding models are often considered feature extractors, which can generate a speech-print (related to finger-print) of an individual. Hence, every individual's speech will remain closer in the embedding space, causing \added{to create} a cluster of embedding\added{s} from speech frames. 

Gaussian mixture model (GMM) supervector \cite{campbell2006support} (stacking mean vectors from GMM), and Joint factor analysis (JFA) \cite{kenny2007joint} have been popularly integrated into the speaker recognition task.  \replaced{JFA }{Joint factor analysis} merges the speaker-dependent and channel-dependent supervector and generates a new supervector based on the dependency. GMM and JFA were significantly accepted as \deleted{a} feature extractor\added{s} and implemented in various speaker recognition strategies. Later on, inspired by JFA, identity vector (i-vector) \cite{dehak2010front} was introduced. I-vector contributes to changing the channel-dependent supervectors\deleted{,} and integrates speaker information within the supervectors. Hence, i-vector became more sensitive to speech variations and greatly accepted by the researchers. In most cases of JFA and i-vector, MFCC is widely implemented. MFCC is a linear cosine transform of a log power spectrum used to extract a sound's information. However, a lower MFCC with a lower cepstral coefficient returns only sound information, whereas the higher value of the coefficient represents speaker information as well \cite{molla2004effectiveness}. Further, probabilistic linear discriminant analysis (PLDA) \cite{ioffe2006probabilistic} is mostly used for implementing speaker verification and identification systems using i-vectors \cite{garcia2011analysis}.

The present improvement of DL architectures has led to revisiting the speech embedding representation neural architecture perspective. Deep vector (d-vector) \cite{variani2014deep} is a mutated implementation of the speech frame embeddings using deep neural networks (DNN).  The d-vector depends on the automated feature extraction process of DNN. The model's training process is supervised, and in the basic implementation of the d-vector, it is explored as a text-dependent system. After the training procedure, the softmax layer is left out, and the embeddings are extracted from the last hidden layer. Although the d-vector is based on DNN, further studies have been made using CNN architectures \cite{chen2015locally}. In the modified architecture, \deleted{the} speech is converted into MEL coefficients, which are normalized and supplied to CNN. Moreover, extensive studies have been made to improve the basic d-vector to a text-independent unsupervised vector generation using domain adaptation \cite{snyder2017deep}. The mechanism is split into two parts in the upgraded version\replaced{: }{,} a DNN that extracts embeddings and a separately trained classifier that classifies speakers. These studies' limitation is that most of them require massive labelled data in the training procedure. Also, the embedding\deleted{s'} performance in the case of unseen speakers dramatically depends on the training data. 

As DNN architectures are dependent on the amount of training data, an improved strategy of the d-vector is proposed, named x-vector \cite{snyder2018x}. X-vectors are a modified version of d-vector, which depends on basic sound augmentation techniques, noise and reverberation. Further, the implementation highly motivates data augmentation usage and presents a decent accuracy improvement over i-vectors. The default x-vector is implemented based on the improved text-independent version of the d-vector \cite{snyder2017deep} by properly utilizing data augmentation.

The present state of the art speech embedding systems tends to be unsupervised. However, the concept of unsupervision still depends on a large set of training data. Both d-vector and x-vectors directly rely on the domain adaptation \cite{redko2019advances} policy of neural network architectures. Hence, the performance of these architectures on unseen data massively depends on the volume and diversity of the training dataset. The domain adaptation capability of neural network architectures is further increased by using synthetic datasets \cite{spyrou2020data}. However, the performance is still dependent on previously learned features\added{,} and performance might lack due to data inefficiency and domain variation \replaced{between }{of} training and testing data.

Therefore, we introduce an approach that is independent of domain adaptation of neural network architectures. Instead, the proposed method tends to utilize the automated feature extraction \added{of} neural network.


%

\begin{figure}[H]
	\centering
	\includegraphics[width=0.65\linewidth]{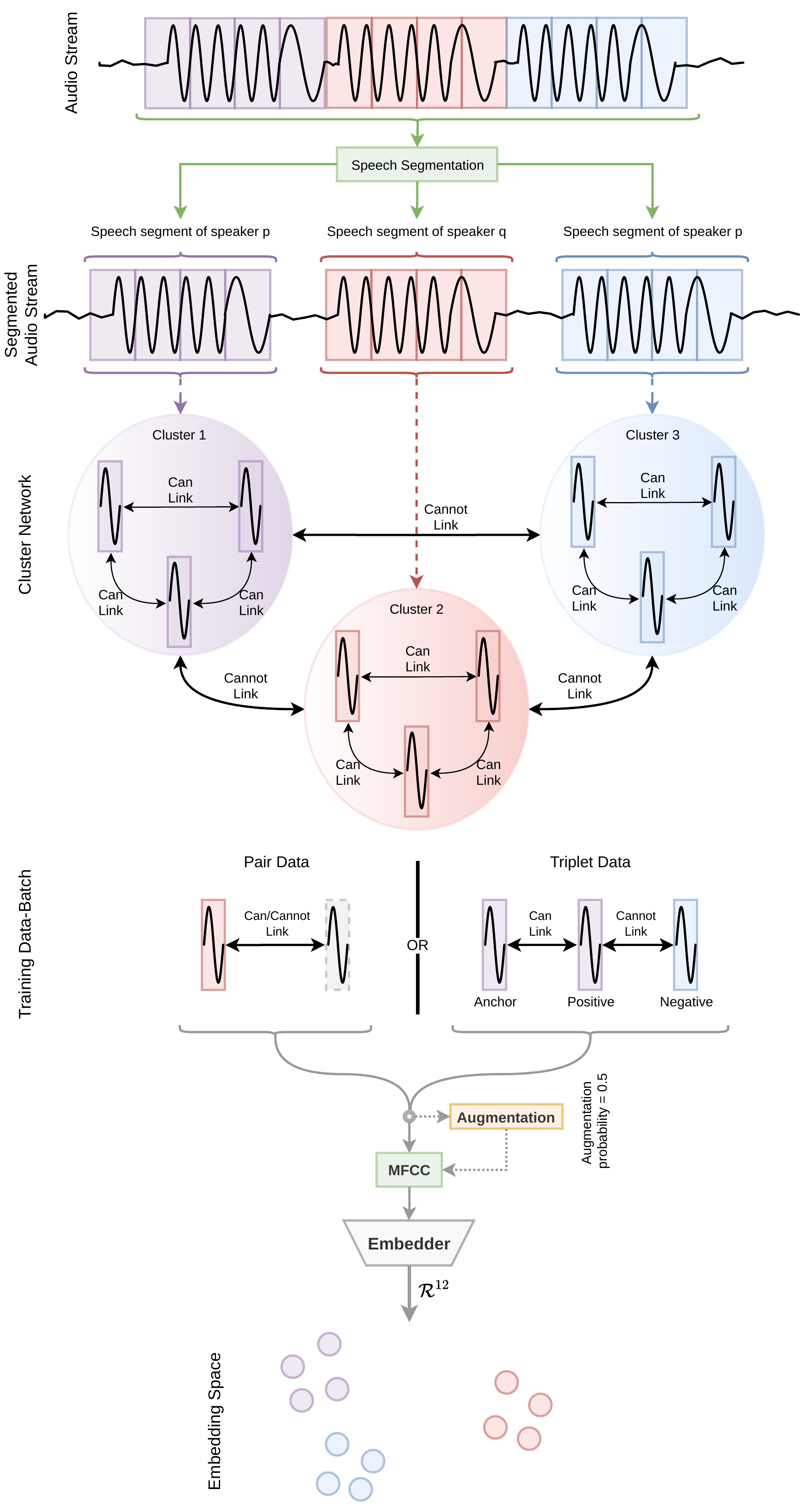}
	\caption{The figure illustrates the comprehensive procedure\deleted{s} to generate u-vectors. Audio stream is segmented and given pseudo labels. Then a cluster network is formed using small non-overlapping speech frames from the speech segments. Finally, training data is generated from the cluster network based on the needs of the siamese architecture. The pairwise network requires an equal number of can-link and cannot link pairs. In contrast, triplet networks receive three data, one pair with can-link, and one cannot-link pair. The network/embedder is trained on randomly augmented speech frames, further converted into 2D feature maps using MFCC. While training, the embedder reforms the cluster associations, where speaker similarity is considered. }
	\label{fig:workflow}
\end{figure}


\section{Methodology}
\label{sec:methodology}

This paper's clusterable speaker embedding generation is based on a particular assumption: a speaker speaks continuously for a specific time. Hence, if segmentation methods are used\deleted{,} or a small speech segment is extracted based on voice segmentation techniques, most speech segments will contain an individual's speech. However, some segments might be impure, i.e., a single segment may have multiple individuals' speech. Nevertheless, we argue that the ratio of impurity would be small enough for most general speech conversations. Hence, such a strategy is investigated with the most common neural network pipelines; a siamese network \cite{guo2017learning}. AutoEmbedder framework \cite{ohi2020autoembedder} is used as a DL architecture to extract speaker embeddings. Figure \ref{fig:workflow} illustrates the basic workflow of producing u-vectors.

In this section, the methodology of generating u-vector is introduced. The section is segmented as follows: the problem formulation and assumptions are defined in Section \ref{sec:formulation}. The proposed work includes speech segments discussed in Section \ref{sec:speech_segment}. Further, speech segments are broken into small speech frames for construction pairwise constraint, \added{which is} addressed in Section \ref{sec:pairwise_constraints}. Uncertainties due to the pseudo-labels of pairwise constraints are discussed in Section \ref{sec:uncertainties}. Challenges of deciding segmentation length are addressed in Section \ref{sec:proof}. Finally, the DL framework AutoEmbedder \cite{ohi2020autoembedder}, which is used to explore the actual cluster linkage, is theorized in Section \ref{sec:ae}.

\subsection{Problem Formulation and Assumptions}
\label{sec:formulation}


The proposed method tends to solve the speaker recognition system in an unsupervised manner based on some constraints. Table \ref{tab:annotations} summarizes the paper's mathematical notations to facilitate the readers. To comprehend the problem statement, let $\mathcal{S}$ be a database of speech segment, where $\mathcal{X}_k$ be a short-length audio segment containing speech of an individual. Also, let $x_i$ be a smaller window/frame of the audio segment, where $x_i \in \mathcal{X}_k$. From a particular speech segment, $\mathcal{X}_k$, $\mathcal{M}$ number of non-overlapping speech frames are generated. As it is stated that a speech segment belongs to a single individual, the smaller speech frames also belong to that individual. From this intuition, we construct pairwise constraints between audio frames. We define two speech frames belong to the same cluster if they belong to the same audio segment. On the contrary, we consider two \replaced{speech frames }{clusters} that belong to different clusters if they belong to different audio segments. Based on the pairwise relations, a set of cluster $\mathcal{C}$ can be generated. Where a single cluster ($c_i \in \mathcal{C}$) belongs to a specific speech segment. Considering most speech segments will belong to a single speaker, we can assume that most cluster $c_i$ would contain a single individual's data. However, as multiple speech segments can belong to a single individual, multiple clusters may contain a single individual's data. Hence, the challenge is to find such optimal cluster relationships such that no two clusters may contain speech of a single individual. The training strategy has two possibilities, such as:
\begin{itemize}
	\item In the case of an unlabeled dataset, let us consider that each audio file contains a single individual's speech. For a set of audio files with an unknown number of speakers, our approach is suitable to produce clusterable embeddings based on speakers. This constraint is similar to semi-supervised learning, as some of the pairwise constraints are known \cite{ren2019semi}.
	\item Let us consider a dataset containing multiple speakers' conversations, where a single audio stream may include various speakers. In such a case, no pairwise constraints are known, and an unsupervised strategy is required. Hence, we produce hypothetical pairwise constraints based on audio segmentation processes such as VAD, word segmentation \cite{kamper2017embedded}, etc., and construct pairwise constraints. However, in such a case, the embedding system's accuracy depends on the purity of the audio segmentation process.
\end{itemize}
For both of the problem, DL architecture is used to aggregate multiple clusters such that the resulting cluster contains all of the embeddings of a single individual. We imply that if a DL function can properly extract speech features from audio frames, it can obtain optimal reasoning of speech frames being similar and dissimilar. Further, an optimally trained DL framework can successfully re-cluster the data based on the feature similarity rather than the number of hypothetical clusters. We extend our investigation towards finding such DL training strategy. 

Since the approach deals with unsupervisely generating speaker embedding vectors from speech data (without domain adaptation), the output embedding vectors are named as unsupervised vectors (u-vectors). Formally u-vector is defined as follows,

\begin{define}
	Unsupervised vectors (u-vectors) refer to a set of DNN generated speech embeddings, which are clusterable based on speakers, trained from unlabeled speech segments.
\end{define}

In the formal definition, by DNN, any specific implementation of substantially deep neural networks are indicated, such as convolutional, feedforward, recurrent, etc.


\begin{table}[H]
	\centering
	\ra{1.2}
	\caption{The mathematical notations used in the paper are summarized.}
	\label{tab:annotations}
	\begin{tabular}{cp{11cm}}
		\toprule
		\textbf{Notation} & \textbf{Description} \\ 
		
		\midrule
		
		$\mathcal{S}$ & A set of audio segments. Audio segments are fragments of a continuous audio stream. We assume that most audio segments contain \deleted{the} speech of a single individual. \\ 
		
		\midrule
		
		$\mathcal{X}$ & A single audio segment, $\mathcal{X} \in \mathcal{S}$. \\ 
		
		\midrule
		
		$x_i$ & An audio frame, generated by taking shorter frames from an audio segment, $x_i \in \mathcal{X}_k$. Audio frames are used to train DL architectures. \\ 
		
		\midrule
		
		$\mathcal{M}$ & Denotes the number of possible audio frames in an audio segment, $x_{1 \le i \le \mathcal{M}} \in \mathcal{X}_k$. Theoretically, $\mathcal{M} \times |x_i| = |\mathcal{X}_k|$. \\ 
		
		\midrule
		
		$\mathcal{C}$ & A set of clusters. These clusters are formed using hypothetical pairwise constraints. As cluster linkages are constructed based on the speech segment relations, it can be considered that $|\mathcal{S}| = |\mathcal{C}|. $ \\ 
		
		\midrule
		
		$c_i$ & Denotes a subset of the entire cluster, $c_i \subseteq \mathcal{C}$. Here, $c_i$ represents a cluster constructed using the inter-relationship of speech frames, belonging to a specific speech segment $\mathcal{X}_i$. \\ 
		
		\midrule
		
		
		$\mathcal{N}$ & The actual number of individuals in $\mathcal{S}$, considering the ground truth. For this specific problem, the value of $\mathcal{N}$ is unknown. \\ 
		
		\midrule
		
		$\alpha$ & The distance hyperparameter used for AutoEmbedder \cite{ohi2020autoembedder} architecture. For other architectures, $\alpha$ may indicate a connectivity state for any two cluster nodes. \\ 
		
		\bottomrule
		
	\end{tabular}
\end{table}

\subsection{Speech Segments}
\label{sec:speech_segment}

To generate the pairwise constraints, it is considered that a speech segment belongs to an individual. Moreover, if it is possible to extract accurate pairwise constraints, a DL framework can be trained \deleted{in} using those constraints. To generate such pairwise constraints, speech segmentation procedures are required.

Speech can be easily segmented using various techniques. Methods such as VAD \cite{tan2020rvad} and word segmentation \cite{brent1999speech} can be indeed adopted to define such speech segments, containing the voice of a single individual. It is also feasible to assume that a single individual mostly speaks more than one word in a conversation. Hence, it is also possible to queue multiple speech segments and hypothesize that they come from a single individual. However, increasing the queue or size of a speech segment also increases the probability of impurity of a speech segment (discussed in Section \ref{sec:proof}). By impurity in a speech segment, it is referred that a speech segment contains more than one speaker. Impure data can often trick the DL frameworks from finding actual relationships among clusters. Hence, to minimize the impurity risk, we study speech segments with a length of one second. After the successful extraction of speech segments, the pairwise constraints are to be constructed. Although the overall framework is dependent on proper speech segmentation techniques, we avoid implementing such segmentation methods. Instead, we provide a detailed evaluation of embedding accuracy based on various levels of cluster impurities.

\subsection{Pairwise Constraints}
\label{sec:pairwise_constraints}

The DL framework is trained based on pairwise constraints. Pairwise constraint contains a pairwise relationship between a pair of inputs.  By considering $x_i$ and $x_j$ as two random speech frames, two circumstances may occur: a) speech frames may belong to the same audio segment $\mathcal{X}_k$ or b) they may belong to different audio segments. In the current state of the problem, as the speech labels' ground truth is unknown for every speech segment, we consider each segment belonging to different individuals. Hence, the number of unique pseudo labels is equal to $|\mathcal{S}|$. Mathematically, $\mathcal{C}$ being a set of clusters, $c_i$ being a particular cluster of similar nodes, and $\mathcal{X}_k$ being a specific speech segment,
\begin{equation}
\begin{split}
\forall x_i \in \mathcal{X}_{k} \;\; 
and \;\;
\forall x_j \in \mathcal{X}_{k}, \;\;\;
x_i, x_j \in c_k \\
\forall x_i \in \mathcal{X}_{k} \;\; 
and \;\;
\forall x_j \notin \mathcal{X}_{k}, \;\; 
x_i, x_j \notin c_k
\end{split}
\label{eq:connection}
\end{equation}

The DL framework is trained based on the defined cluster constraints. To properly introduce the inter-cluster and intra-cluster relation to a DL framework, we define a gound regression function based on pairwise criteria derived in Eq. \ref{eq:connection}. The function is defined as,
\begin{equation}
\label{eq:pairwise}
\mathcal{P}_c(x_i, x_j) = 
\begin{cases}
0 & \text{if $x_i, x_j \in c_p$} \\
\alpha  &  \text{if $x_i \in c_p$ and $x_j \in c_q$ }
\end{cases}
\end{equation}

In general, the $\mathcal{P}_c(\cdot, \cdot)$ outputs the distance constraints that each embedding (generated from speech frames) holds. The function infers that an embedding pair must be at a close distance if they belong to the same cluster or at a distance of $\alpha$ otherwise. However, embedding pairs belonging to different clusters may be at a distance greater than $\alpha$, which is established in the AutoEmbedder architecture (Eq. \ref{eq:relu}). We use the pairwise constraints to train a DL architecture. Further, we revisit the data-clusters' uncertainty and segmenting impurities and explore why a DL framework may be necessary in such a case.



\subsection{Uncertainties in Pairwise Constraints}
\label{sec:uncertainties}


The cluster assignments are mostly uncertain based on two major concerns: a) the segmented audio $\mathcal{X}_k$ may be impure, b) the ground-truth of cluster assignments are unknown. Therefore, in most cases, the number of ground-label (defined as $\mathcal{N}$) is theoretically not equal to the number of clusters, i.e., $\mathcal{N} \neq |\mathcal{C}|$ and $\mathcal{N} \neq |\mathcal{S}|$, where $|\mathcal{S}| = |\mathcal{C}|$. Moreover, due to such impurity and uncertainty of ground-labels, the subsequent flaws in the training dataset (based on pairwise properties) are frequently observed,
\begin{itemize}
	\item \textbf{Impurity in must-link constraint:} The dataset's core concept is to assume that an audio segment $\mathcal{X}_k$ contains only one individual's speech. Generally, a segmentation system may inaccurately identify speech segments and hold multiple individuals' speech in a single audio segment. However, if we consider short length audio segments, the probability of speaker fusion rapidly decreases.
	
	\item \textbf{Error in cannot-link constraint:} Let, $x_i \in c_p$ and $x_j \in c_q$, where $c_p \neq c_q$. The cluster assignments are considered based on the number of audio segments. Hence, for most datasets, the number of speech segments is greater than the actual number of speakers, $|\mathcal{C}| \ge \mathcal{N}$. Therefore, considering the ground-truth, the assumption $c_p \neq c_q$ may be wrong, and data pair $x_i$ and $x_j$ may belong to the same cluster considering the ground truth.

\end{itemize}

If we consider a  cluster network $\mathcal{C}$ with no impurity, then the task of DL is to eliminate the errors in cannot link constraints based on the feature relationship. Hence, if it is possible to prioritize the speech features to a DL framework, it can allegedly aggregate appropriate cluster from erroneous cannot-link clusters. Therefore, training the DL architecture reduces errors in cannot-link constraints. However, reducing the impurity of the input data's must-link constraints considerably depends on the length of speech segments and segmentation policies.


\subsection{Segment Length Analysis}
\label{sec:proof}

The time-domain length of the speech segments (defined by $|\mathcal{X}|$) operates a vital role in the overall performance of the training process. Each segment is further windowed into smaller speech frames. Hence, the segment length must be divisible by the length of fixed-size speech frames (defined by $|x|$). Various architectures consider overlapped frames while windowing speech signals. However, we avoid such measures, as such overlaps result in mixing similar speech patterns in multiple speech frames.

To illustrate the trade-off of selecting an optimal length of speech segment $|\mathcal{X}|$, let us consider $\mathcal{L}_{mean}$ being the mean and $\mathcal{L}_{std}$ the standard deviation (std) of the length of speech segmentations for a given dataset (or a buffer of audio stream). Therefore, statistically, $\mathcal{L}_{mean} - \mathcal{L}_{std}$ is the optimal minimal length for which we can assume that most segments strictly contains speech of a single individual. However, if the minimum segment length is considered, the number of frames per segment $\mathcal{M}$ would also reduce. 

Reducing the number of frames per segment due to a shorter segment would deliver less inter-cluster relations for each segment. The reduction of inter-cluster association would also cause the DL framework struggle finding feature relation between speech frames. Further, increasing the size of speech segments may also result in impure components, if  $|\mathcal{X}| \ge \mathcal{L}_{mean} - \mathcal{L}_{std}$.

To explore the reason of impurity, let us consider an audio stream contains a mean time $\mathcal{J}_{mean}$ with standard deviation of $\mathcal{J}_{std}$, after which, the speaker exchanges. In such condition, selecting the length of segment too high may result in being $|\mathcal{X}| \ge \mathcal{J}_{mean}-\mathcal{J}_{std}$. However, statistically, in most general conversations, the length of minimal speech segmentation is mostly less than the speaker exchange time, $\mathcal{L}_{mean} + \mathcal{L}_{std} \le \mathcal{J}_{mean}-\mathcal{J}_{std}$. Therefore, if we can select such $|X|$, for which, $|\mathcal{X}| < \mathcal{L}_{mean} - \mathcal{L}_{std}< \mathcal{J}_{mean}-\mathcal{J}_{std}$, the rate of impurity would be zero. Hence, selecting $|\mathcal{X}| \approx \mathcal{L}_{mean}$ would reduce the rate of impurity. For the experimental datasets, the $\mathcal{L}_{mean}$ is equal to one (illustrated in Table \ref{tab:dataset}). Hence, we experiment with one-second speech segment. Further, we investigate a pairwise framework in which we try to trick DL architecture into converging towards the ground cluster relationship.

\subsection{AutoEmbedder Architecture}
\label{sec:ae}

As a DL architecture, we use a pairwise constraint-based AutoEmbedder framework to re-cluster speech data. However,  we introduce further modifications to the network's general training process to strengthen the learning progress. In general, the AutoEmbedder architecture is trained based on the pairwise constraints defined by function $\mathcal{P}_c(\cdot,\cdot)$. The architecture follows siamese network constraints that can be presented as,
\begin{equation}
\label{eq:snn}
\mathcal{S}(x, x^{'}) = ReLU(\|\mathcal{E}_{\phi}(x) - \mathcal{E}_{\phi}(x^{'})\|, \alpha) = \mathbb{R}_{\leq \alpha}^{+}
\end{equation}

The $ReLU(\cdot, \cdot)$ function used in Eq. \ref{eq:snn} is a thresholded ReLU function, such that,
\begin{equation}
\label{eq:relu}
ReLU(x, \alpha) = 
\begin{cases}
x & \text{if $0 \leq x < \alpha$} \\
\alpha  &  \text{if $x \geq \alpha$}
\end{cases}
\end{equation}
In Eq \ref{eq:snn}, the $\mathcal{S}(\cdot, \cdot)$ represents a siamese network function, that receives two inputs. The framework contains a single shared DNN network  $\mathcal{E}_{\phi}(\cdot)$ that map higher dimensional input to a lower dimension clusterable embeddings. The Euclidean distance of the embedding pairs is calculated and passed through the thresholded ReLU activation function derived in Eq \ref{eq:relu}. The threshold value is a cluster margin of $\alpha$. Due to the threshold, the siamese architecture always generates outputs in the range $[0, \alpha]$. The L2 loss function is used to train the general AutoEmbedder architecture. The AutoEmbedder architecture is trained using an equal number of must-link and cannot link constraints for each data batch. However, in a triplet architecture, the problem is automatically solved, as each triplet contains a fusion of cannot link (negative) and can-link (positive) data.

\subsection{Augmenting Training Data}

Both types of cluster relationships (can-link and cannot-link) may contain faulty assumptions and pseudo labels considering the ground truth. Hence, a basic augmentation scheme is used to trick the DL network from overfitting erroneous cluster relationships. Although various augmentation techniques are available, we adhere to mixing noise with speech data for augmentation. For noise augmentation, we implement a basic formula that is,
\begin{equation}
\label{eq:augment}
Aug(x_i, noise, thres) = \{x_{i}^{'} | x_{i}^{'} = x_{i} \times (1-thres) + noise \times thres \} \;\;\; [0 \le thres \le 1] \
\end{equation}
Here, $Aug(\cdot,\cdot,\cdot)$ is a function that produces augmented speech data, which is inputted as $x_i$. The $thres$ is a threshold used to define the ratio of mixing $noise$ with speech data $x_i$. Augmenting noise with speech frames results in less-confusing the AutoEmbedder network in case of erroneous data pairs. Fusing noise may facilitate the architecture by ignoring faulty data pairs due to different noise situations. Moreover, augmenting data also results in data variation, and the network extracts more beneficial features from speech data.  Algorithm \ref{alg:algo} presents pseudocode of the pairwise training process.

\begin{algorithm*}
	\KwIn{
		Dataset $D$ containing speech frames,
		DL model with initial weights $\mathcal{E}_{\phi}$,
		Distance hyperparameter $\alpha$,
		Training epochs $E_p$
	}
	
	\BlankLine
	
	Initialize siamese network, $\mathcal{S}_{\phi}(\cdot,\cdot) \gets ReLU(||\mathcal{E}_{\phi}(\cdot) - \mathcal{E}_{\phi}(\cdot)||, \alpha)$ \\
	
	\BlankLine
	
	\For{$epoch \gets 1$ \KwTo $E_p$} {
		\ForEach{$\mathcal{D}_{batch} \in \mathcal{D}$}{
			$\mathcal{X}, \: \mathcal{X}^{'}, \: \mathcal{Y} \gets \{\}, \: \{\}, \: \{\}$ \\
			$counter \gets 0$\\
			
			\ForEach{$x \in \mathcal{D}_{batch}$}{
				$\mathcal{X}$ $\gets$ append $x$ in $\mathcal{X}$ \\ 
				
				\If{$counter < |batch|/2$} {	
					$\mathcal{X}^{'} \gets$ randomly pick and append a can-link speech frame from $\mathcal{D}$ \\
					$\mathcal{Y} \gets$ append $0$ in $\mathcal{Y}$ \\
				}
				\Else {
					$\mathcal{X}^{'} \gets$ randomly pick and append a cannot-link speech frame from $\mathcal{D}$ \\
					$\mathcal{Y} \gets$ append $\alpha$ in $\mathcal{Y}$
				}
				
				
				$counter \gets counter + 1$\\
			}	
			\BlankLine
			$\mathcal{X} \gets$ randomly select half of the speech frames and augment them \\
			
			$\mathcal{X}^{'} \gets$ randomly select half of the speech frames and augment them \\
			
			$S_{\phi} \gets$ Train $S_{\phi}$ with $\mathcal{X}, \mathcal{X^{'}}, \mathcal{Y}$ \\
			
	}}
	\caption{AutoEmbedder training for speaker recognition.}
	\label{alg:algo}
\end{algorithm*}



%
%

\section{Experiments and Evaluations}
\label{sec:experiment}

In this section, the proposed scheme is experimented based on the impurity of speech segmentation. As the architecture's target is to produce clusterable embedding, we use k-means to measure the purity of the clusters generated by the embedding system. Further,  three popular metrics, Accuracy (ACC), normalized mutual information (NMI), and adjusted rand index (ARI), are used to measure clustering effectiveness. The metrics are calculated as demonstrated in \cite{ohi2020autoembedder}, and are widely implemented to refer to the purity of clustering \cite{ren2019semi,yang2019deep}.


\subsection{Experimental Setup}
\label{sec:setup}

\deleted{Tensorflow [31] and Keras [32] are used to implement the neural network architectures. Moreover, scikit-learn [33] is used to implement clustering algorithms. Numpy [34] is used to perform efficient numerical operations. Librosa [35] is used to perform audio analysis.} The datasets were segmented using a threshold of 16 decibels, implemented as a VAD. The audio streams have been processed with a sample rate of 16000Hz. For audio to spectrogram conversion, the parameters are set as described, size of fast-fourier transform: 191, window-size: 128, stride: 34, mel-scales: 100. Speech spectrograms are used as inputs to train the DL architectures.

\subsection{Datasets}
\label{sec:datasets}

For experimentation, three speech datasets \replaced{have }{has} been used. TIMIT \cite{garofolo1993darpa} and LibriSpeech \cite{panayotov2015librispeech} \replaced{are }{is} popular speech datasets for English language. Moreover, we use Bengali Automated Speech Recognition Dataset \cite{kjartansson-etal-sltu2018} to show the diversity of our approach for additional languages. Among the three datasets, TIMIT and LirbriSpeech datasets contain studio-grade audio speech. Bengali ASR dataset is crowdsourced hence, contains a diverse sound and noise variation. Table \ref{tab:dataset} illustrates some basic statistics of each dataset. Throughout the experiment, we abbreviate LibriSpeech and Bengali ASR dataset as LIBRI and ASR, respectively. As the training procedure augments noise with the speech frames, we use scalable noisy speech dataset \cite{reddy2019scalable}. The dataset contains diverse environmental noises, which helps the architectures to explore and relate speech features.

%

\begin{table}[H]
	\centering
	\ra{1.0}
	\caption{The table illustrates the mean, median and deviation of segment duration and words per sentence for each dataset. The segment duration is calculated using the setup described in Section \ref{sec:setup}.\\}
	\label{tab:dataset}
	\hspace{0.2pt}
	\begin{adjustbox}{max width=\textwidth}
		\begin{tabular}{@{}crrrcrrr@{}}\toprule
			\multirow{2}{*}{Dataset} & \multicolumn{3}{c}{Segment duration} & \phantom{ab}& \multicolumn{3}{c}{Words per sentence}\\
			
			\cmidrule{2-4} \cmidrule{6-8}
			& Mean & Median & STD && Mean & Median & STD \\ \midrule
			
			{TIMIT \cite{garofolo1993darpa}} & 1 & 0.8 & 0.6 && 8.63 & 8.0 & 2.6\\ \midrule
			
			{LibriSpeech \cite{panayotov2015librispeech}} & 1.2 & 0.8 & 1 && 18.9 & 15.0 & 12.9\\ \midrule
			
			{Bengali ASR \cite{kjartansson-etal-sltu2018}} & 1.3 & 1.2 & 0.8 && 3.20 & 3.0 & 3.0\\ 
			
			\bottomrule
		\end{tabular}
	\end{adjustbox}
\end{table}

\subsection{Result Analysis}
\label{sec:result_analysis}

Two methods are implemented, AutoEmbedder (pairwise architecture) and a triplet architecture, to analyze the speech embeddings based on the proposed strategy. However, apart from these two strategies, the currently famous speech vector methods do not hold to the training properties considered in the paper. They mostly follow a supervised learning or domain adaption strategy. Hence, they are disregarded in this experiment.

\begin{figure}[H]
	\centering
	\includegraphics[width=1\linewidth]{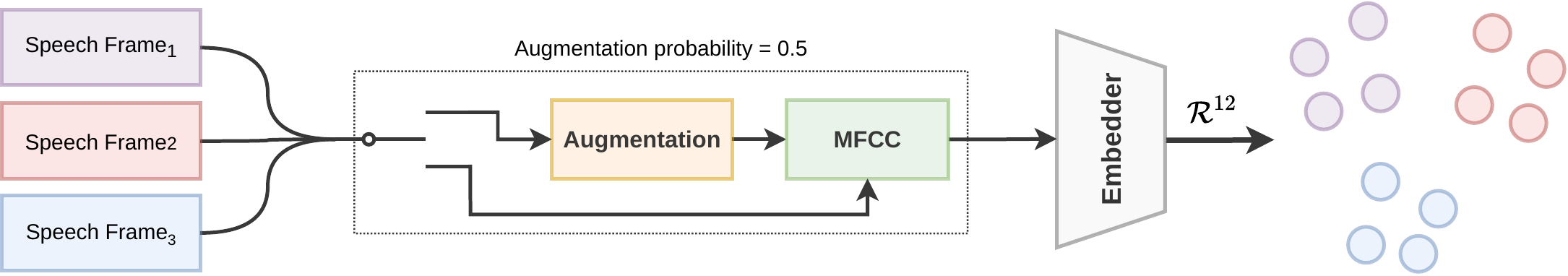}
	\caption{The training of both pairwise and triplet networks is done through the same data processing pipeline. Half of the inputs are randomly augmented and processed using MFCC. The MFCC of each frame is further passed to the DL frameworks.}
	\label{fig:imp}
\end{figure}

DenseNet121 \cite{huang2017densely}, is used as a baseline architecture for both of the DL frameworks. Further, both models are connected with a dense layer consisting of $12$ nodes. Therefore, both pairwise and triplet networks produce $12$ dimension embedding vectors. L2-normalization is added on the output layer of the triplet network, as it is suggested that it increases the accuracy of the framework \cite{hermans2017defense}. For AutoEmbedder architecture, the default l2-loss is implemented, whereas the triplet architecture is trained using semi-hard triplet loss. The training pipeline is illustrated in Figure \ref{fig:imp}.

\begin{figure}[h]
	\centering
	\includegraphics[width=0.9\linewidth]{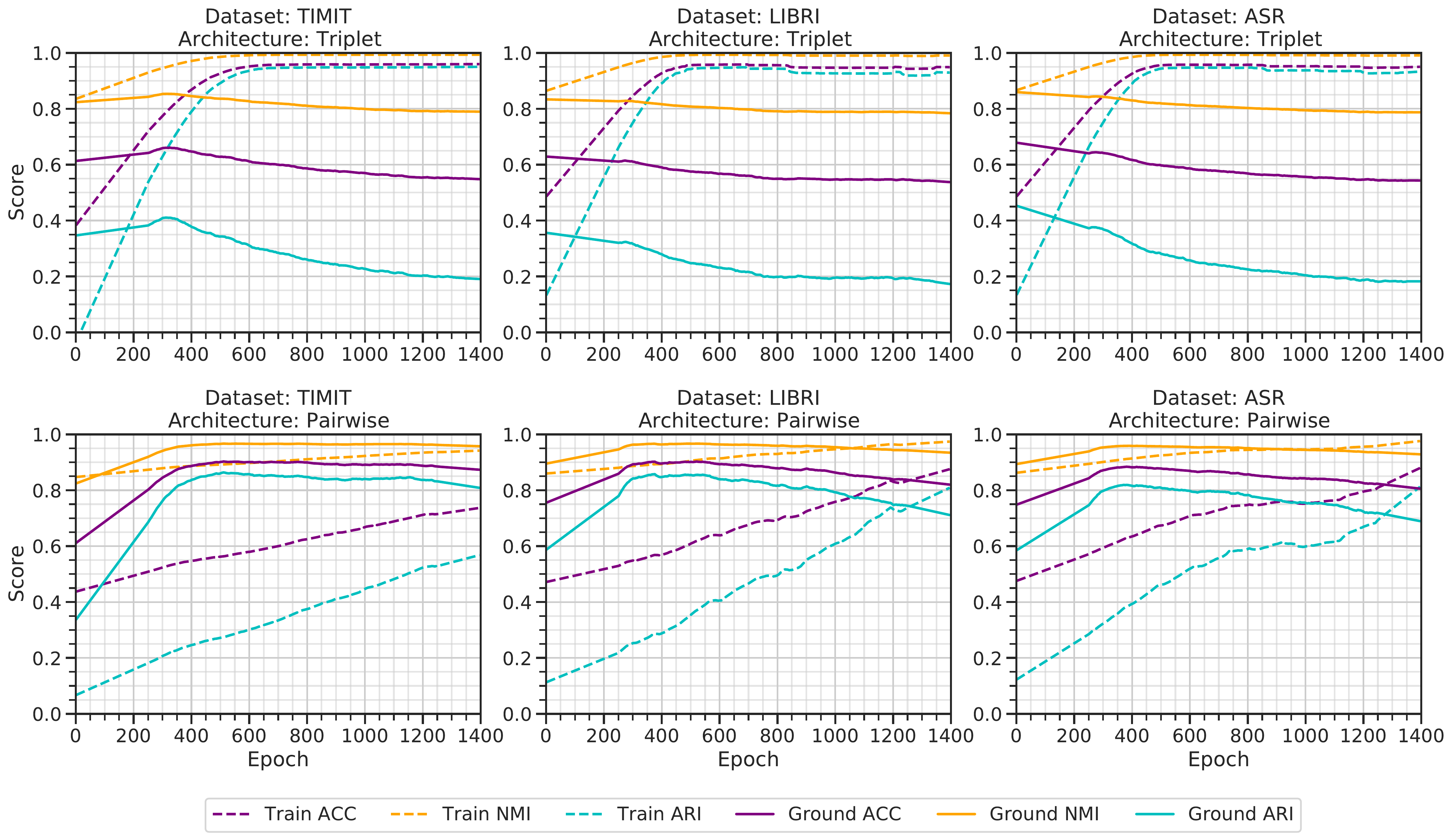}
	\caption{The graphs illustrated in the first row visualizes the metrics on training and ground dataset containing 25 speakers with $impurity=0$ for triplet architecture. The lower row envisions the same for the pairwise architecture. Each column represents benchmarks carried on a single dataset.}
	\label{fig:comp_25spk_0imp}
\end{figure}

\begin{figure}[h]
	\centering
	\includegraphics[width=0.9\linewidth]{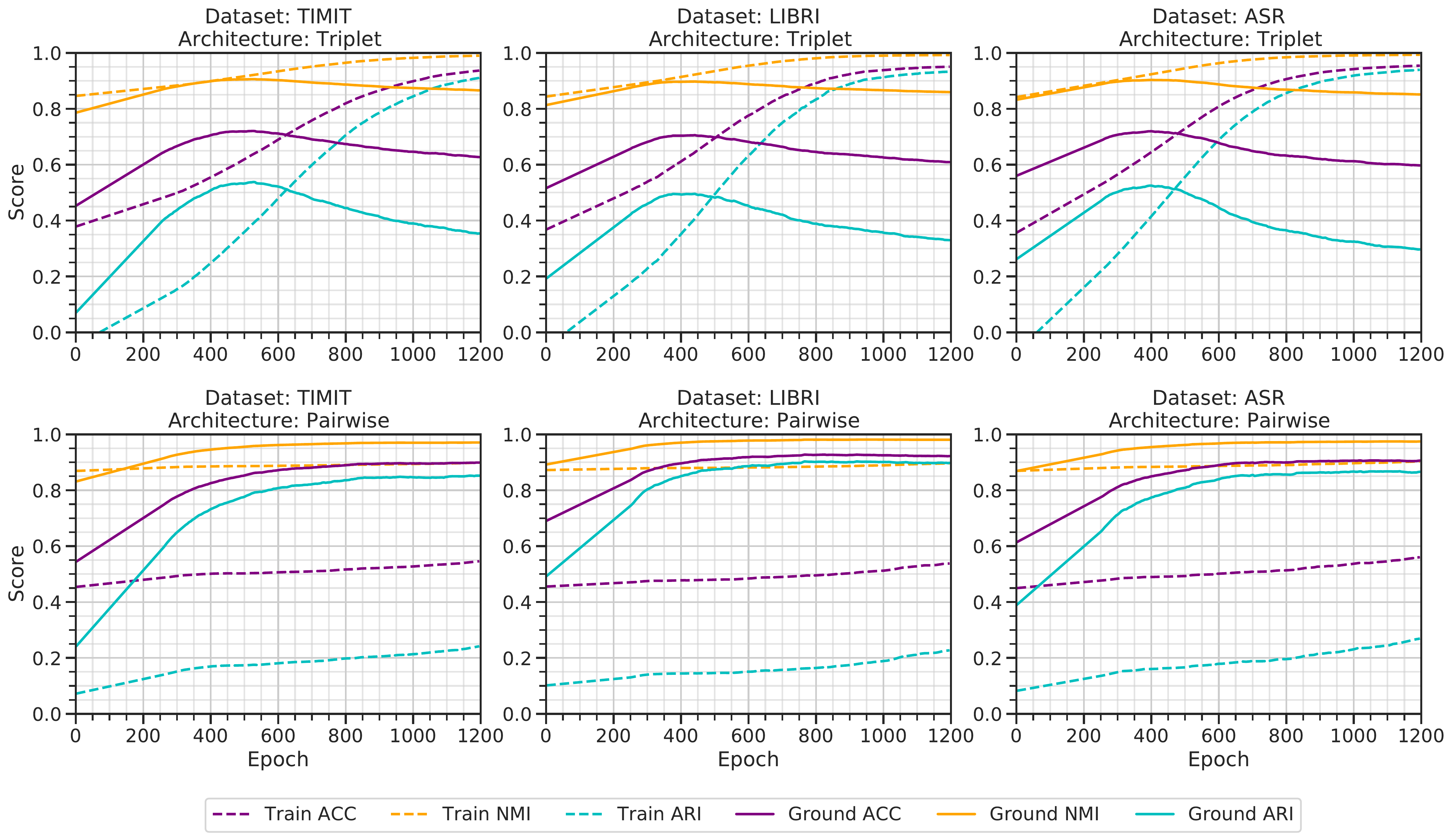}
	\caption{The graph illustrates metrics on training and ground dataset containing 50 speakers with $impurity=0$.}
	\label{fig:comp_50spk_0imp}
\end{figure}

The evaluation process guarantees that both architectures are trained using the same dataset/data-subset. As the training process is unsupervised, the architectures receive the same data for the training and testing process. However, for the training process, the labels are unknown and generated based on the paper's assumptions. We refer to such a dataset as a training dataset. By ground dataset, we refer to the same dataset that further considers the ground truth values. For training both frameworks, we used a batch size of $128$. The training is conducted using Adam \cite{kingma2014adam} optimizer with a learning rate of $0.0005$. \added{The learning rate and batch size were determined using a grid search. Although batch size 64 and 128 result in better evaluation, batch size 128 was considered due to faster computation. }

The training phase's data processing includes heavy computational complexity, including online noise augmentation and spectrogram conversion. Each dataset is randomly augmented with a threshold range of $[0, 0.07]$ for the augmentation process. Further, computing ACC, NMI, and ARI metrics require quadratic time complexity. Hence, we limit the number of speakers to 150. Instead of training on the overall dataset, we train on a sub-set of data, where each speaker contains a speech of 10 seconds. For testing the ground truth data, a random selection of 2-second speech is selected for each speaker. To inquire the architectures properly, we scramble the training dataset's pseudo labels, that produces a can-link impurity in the dataset labels. Hence, we use the impurity ratio as a training data situation to illustrate the rate of impure cluster assignments on the training data pseudo-labels. \added{The models were kept training until the performance on ground truth labels did not improve within the previous 1500 epochs.}

\begin{figure}[h]
	\centering
	\includegraphics[width=0.9\linewidth]{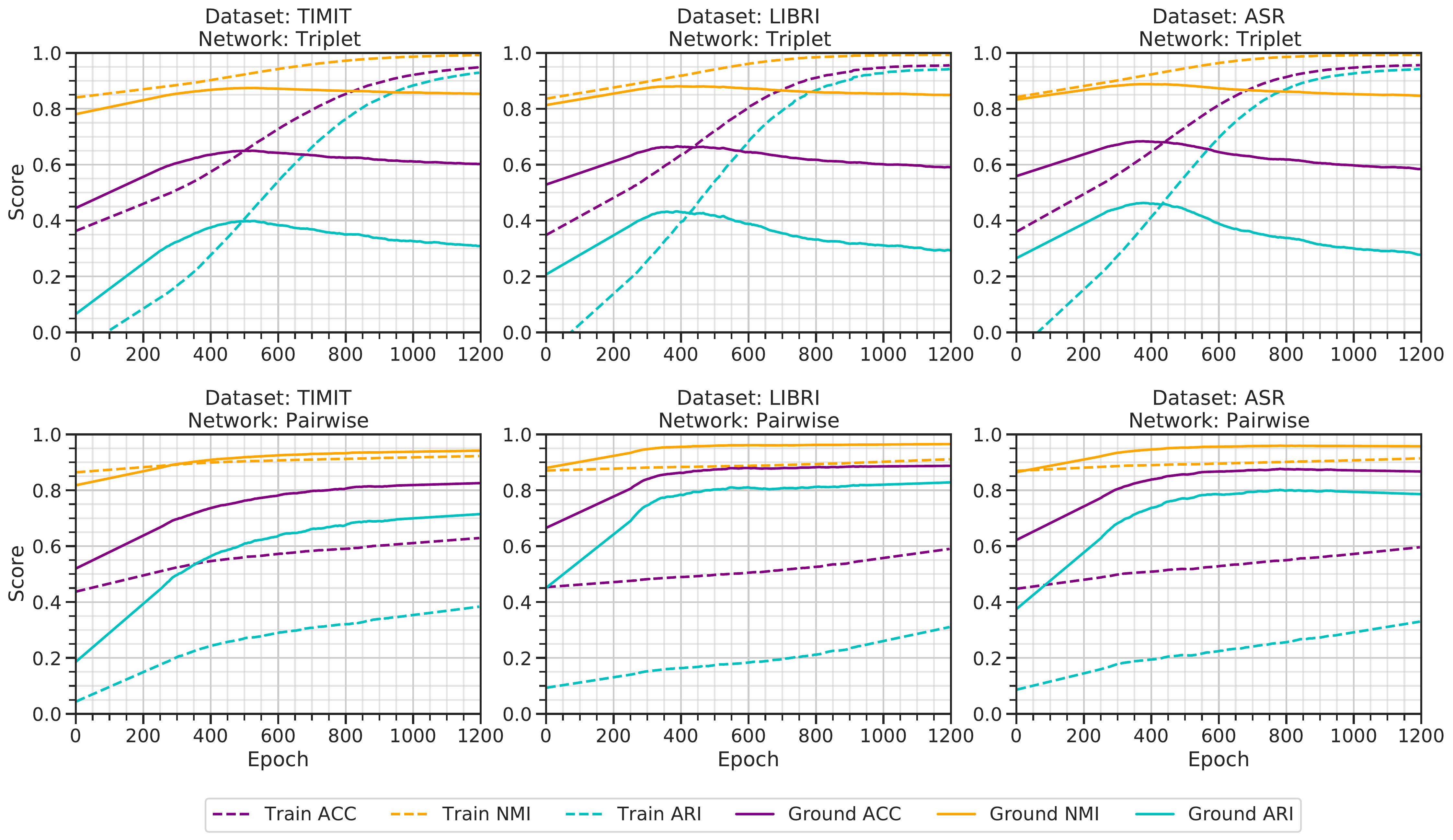}
	\caption{The graph illustrates metrics on training and ground dataset containing 50 speakers with $impurity=0.05$.}
	\label{fig:comp_50spk_0.05imp}
\end{figure}

\begin{figure}[h]
	\centering
	\includegraphics[width=0.9\linewidth]{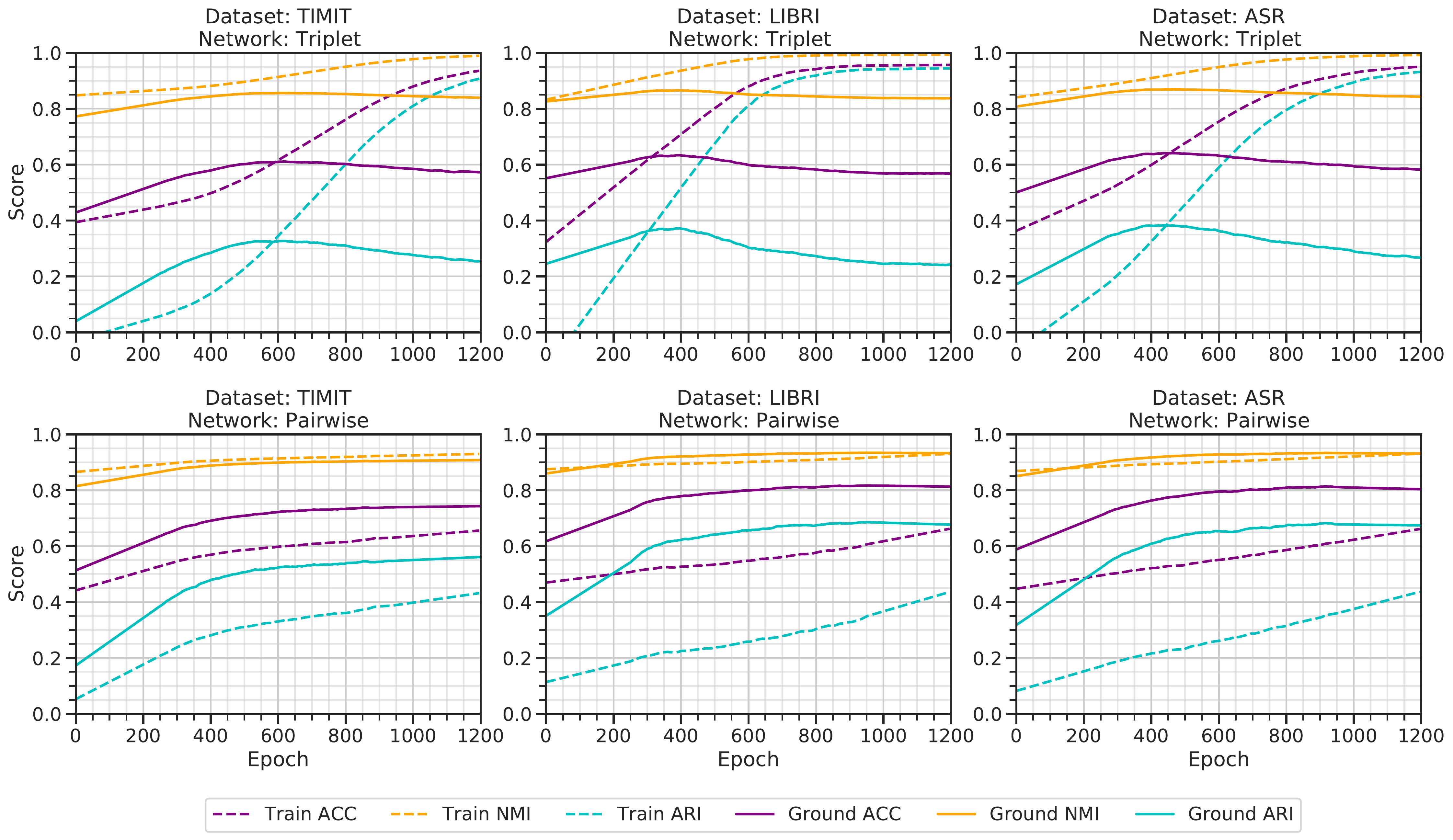}
	\caption{The graph illustrates metrics on training and ground dataset containing 50 speakers with $impurity=0.1$.}
	\label{fig:comp_50spk_0.1imp}
\end{figure}

Figure \ref{fig:comp_25spk_0imp} illustrates a benchmark of the triplet and pairwise architectures while training on three different datasets, with $speakers=25$ and $impurity=0$. The triplet architectures smoothly learn from the training data and greatly overfits on the augmented training data. The ground dataset benchmark is also as expected since the triplet architectures' correctness first increases and gradually decreases due to overfitting. Hence, from the visualization, it can be acknowledged that the triplet network only memorizes the speech features concerning the pseudo labels assigned to them.

\begin{table}[h]
	\centering
	\ra{1.0}
	\caption{The table benchmarks the pairwise architecture in TIMIT dataset with four groups of speakers, 25, 50, 100, and 150. For each group of speakers, the table also considers three segmentation impurities, 0, 0.05, and 0.1 to illustrate the shortcomings of incorrect segmentation, for fully unsupervised speaker recognition strategy.}
	\label{tab:timit}
	\hspace{0.2pt}
	\begin{adjustbox}{max width=0.9\textwidth}
		\begin{tabular}{@{}rrrrcrrrcrrr@{}}\toprule
			& \multicolumn{3}{c}{$Impurity = 0$} & \phantom{ab}& \multicolumn{3}{c}{$Impurity = 0.05$} &
			\phantom{ab} & \multicolumn{3}{c}{$Impurity = 0.1$}\\
			
			\cmidrule{2-4} \cmidrule{6-8} \cmidrule{10-12}
			& $ACC$ & $NMI$ & $ARI$ && $ACC$ & $NMI$ & $ARI$ && $ACC$ & $NMI$ & $ARI$\\ 
			
			\midrule
			
			$Speakers=25$\\
			
			$Train$ & 0.772 & 0.950 & 0.627 && 0.868 & 0.974 & 0.795 && 0.893 & 0.980 & 0.842 \\
			
			$Ground$ & 0.934 & 0.979 & 0.919 && 0.886 & 0.958 & 0.822 && 0.777 & 0.924 & 0.630 \\ \midrule
			
			$Speakers=50$\\
			
			$Train$ & 0.663 & 0.930 & 0.436 && 0.659 & 0.930 & 0.435 && 0.678 & 0.935 & 0.464 \\
			
			$Ground$ & 0.924 & 0.979 & 0.898 && 0.849 & 0.948 & 0.755 && 0.764 & 0.915 & 0.623 \\ \midrule
			
			$Speakers=100$\\
			
			$Train$ & 0.509 & 0.899 & 0.175 && 0.523 & 0.904 & 0.208 && 0.590 & 0.920 & 0.305 \\
			
			$Ground$ & 0.889 & 0.973 & 0.847 && 0.809 & 0.948 & 0.723 && 0.719 & 0.914 & 0.544 \\ \midrule
			
			$Speakers=150$\\
			
			$Train$ & 0.471 & 0.861 & 0.117 && 0.500 & 0.902 & 0.155 && 0.519 & 0.907 & 0.194 \\
			
			$Ground$ & 0.815 & 0.955 & 0.734 && 0.718 & 0.924 & 0.645 && 0.620 & 0.896 & 0.480 \\
			
			\bottomrule
		\end{tabular}
	\end{adjustbox}
\end{table}

On the contrary, the pairwise architecture generates a satisfactory performance, with some irregularities. In general, deep learning architectures produce higher accuracy on the training dataset than validation dataset. However, in such a case, the ground dataset's performance is mostly more elevated than the training dataset. Yet, the performance on the ground datasets generally decreases after 400 epochs. As the number of speakers is small, the architecture easily gets overfitted on the training dataset. Further increasing the number of speakers to 50 reduces the overfitting on training data, as illustrated in Figure \ref{fig:comp_50spk_0imp}. The triplet architecture still overfits on the training data's pseudo label, whereas, the pairwise architecture gives a balanced performance on the ground dataset. 

Increasing the impurity of the inter-connection of the training data reduces the performance of the architectures. \deleted{In} Figure \ref{fig:comp_50spk_0.05imp} and \ref{fig:comp_50spk_0.1imp} illustrates benchmarks conducted with $impurity=0.05$ and $impurity=0.1$ while considering $speakers=50$. The triplet architecture still overfits on the training architecture. In contrast, pairwise architecture slowly memorizes the training dataset. Yet, it holds a marginal exactness on the ground data before overfitting on the training data.

\begin{table}[h]
	\centering
	\ra{1.0}
	\caption{The table benchmarks the pairwise architecture in LIBRI dataset with four groups of speakers, 25, 50, 100, and 150. For each group of speakers, the table also considers three segmentation impurities, 0, 0.05, and 0.1 to illustrate the shortcomings of incorrect segmentation, for fully unsupervised speaker recognition strategy.}
	\label{tab:libri}
	\hspace{0.2pt}
	
	\begin{adjustbox}{max width=0.9\textwidth}
		\begin{tabular}{@{}rrrrcrrrcrrr@{}}\toprule
			& \multicolumn{3}{c}{$Impurity = 0$} & \phantom{ab}& \multicolumn{3}{c}{$Impurity = 0.05$} &
			\phantom{ab} & \multicolumn{3}{c}{$Impurity = 0.1$}\\
			
			\cmidrule{2-4} \cmidrule{6-8} \cmidrule{10-12}
			& $ACC$ & $NMI$ & $ARI$ && $ACC$ & $NMI$ & $ARI$ && $ACC$ & $NMI$ & $ARI$\\ 
			
			\midrule
			
			$Speakers=25$\\
			
			$Train$ & 0.936 & 0.989 & 0.911 && 0.954 & 0.992 & 0.940 && 0.956 & 0.993 & 0.943 \\
			
			$Ground$ & 0.946 & 0.983 & 0.935 && 0.866 & 0.948 & 0.808 && 0.806 & 0.918 & 0.673 \\ \midrule
			
			$Speakers=50$\\
			
			$Train$ & 0.730 & 0.947 & 0.555 && 0.741 & 0.950 & 0.576 && 0.755 & 0.953 & 0.594 \\
			
			$Ground$ & 0.951 & 0.989 & 0.933 && 0.906 & 0.971 & 0.857 && 0.863 & 0.957 & 0.778 \\ \midrule
			
			$Speakers=100$\\
			
			$Train$ & 0.482 & 0.891 & 0.133 && 0.511 & 0.900 & 0.186 && 0.588 & 0.920 & 0.309 \\
			
			$Ground$ & 0.924 & 0.984 & 0.921 && 0.885 & 0.970 & 0.831 && 0.840 & 0.949 & 0.700 \\ \midrule
			
			$Speakers=146$\\
			
			$Train$ & 0.490 & 0.780 & 0.109 && 0.493 & 0.900 & 0.150 && 0.515 & 0.906 & 0.190 \\
			
			$Ground$ & 0.932 & 0.987 & 0.916 && 0.797 & 0.949 & 0.678 && 0.713 & 0.923 & 0.607 \\
			
			\bottomrule
		\end{tabular}
	\end{adjustbox}
\end{table}

Triplet architecture, trained over semi-hard triplet loss strictly targets the positive points (based on the anchor) and minimizes the distance between the anchor and positive data. For negative points, the loss function also strictly distance the embeddings. As the loss function heavily maintains the criteria mentioned above, the architecture gets overfitted on the hypothetical constrains disregarding the actual feature-dependent relations.

In contrast, AutoEmbedder architecture learns to extract features rather than being overfitted in the training data. The reasoning lies in the training strategy of the network. The l2-loss does not strictly consider learning the hypothetical constraints and learns for each batch of data aggregately. As the architecture is not strictly supervised using the loss function, it obtains the feature similarity of audio data. Therefore the architecture re-clusters the data on hyperspace based on feature similarity. 

We further investigate with AutoEmbedder based pairwise architecture with various speaker and impurity condition. Table \ref{tab:timit}, \ref{tab:libri}, and \ref{tab:asr} illustrates the metrics on training and ground dataset for TIMIT, LIBRI and, ASR datasets, respectively. The table illustrates a detailed overview of the performance variation based on the number of speakers and impurity in the training dataset. The pairwise architecture maintains a marginal performance with $impurity=0$ on every dataset. However, increasing the number of speakers results in reducing the performance of the architecture. On the contrary, increasing the impurity of the speech segment further reduces the performance of the architecture. A small fluctuation is observed for LIBRI and ASR dataset while the number of speakers is kept on 25 and 50. Increasing the number of speakers from 25 to 50 causes an increase in accuracy, which is inconsistent. 

\begin{table}[h]
	\centering
	\ra{1.0}
	\caption{The table benchmarks the pairwise architecture in ASR dataset with four groups of speakers, 25, 50, 100, and 150. For each group of speakers, the table also considers three segmentation impurities, 0, 0.05, and 0.1 to illustrate the shortcomings of incorrect segmentation, for fully unsupervised speaker recognition strategy.}
	\label{tab:asr}
	\hspace{0.2pt}
	
	\begin{adjustbox}{max width=0.9\textwidth}
		\begin{tabular}{@{}rrrrcrrrcrrr@{}}\toprule
			& \multicolumn{3}{c}{$Impurity = 0$} & \phantom{ab}& \multicolumn{3}{c}{$Impurity = 0.05$} &
			\phantom{ab} & \multicolumn{3}{c}{$Impurity = 0.1$}\\
			
			\cmidrule{2-4} \cmidrule{6-8} \cmidrule{10-12}
			& $ACC$ & $NMI$ & $ARI$ && $ACC$ & $NMI$ & $ARI$ && $ACC$ & $NMI$ & $ARI$\\ 
			
			\midrule
			
			$Speakers=25$\\
			
			$Train$ & 0.920 & 0.985 & 0.882 && 0.945 & 0.990 & 0.923 && 0.954 & 0.992 & 0.941 \\
			
			$Ground$ & 0.909 & 0.969 & 0.869 && 0.854 & 0.941 & 0.768 && 0.789 & 0.907 & 0.643 \\ \midrule
			
			$Speakers=50$\\
			
			$Train$ & 0.771 & 0.955 & 0.617 && 0.614 & 0.918 & 0.355 && 0.666 & 0.933 & 0.448 \\
			
			$Ground$ & 0.930 & 0.983 & 0.912 && 0.903 & 0.966 & 0.835 && 0.841 & 0.948 & 0.751 \\ \midrule
			
			$Speakers=100$\\
			
			$Train$ & 0.506 & 0.899 & 0.179 && 0.536 & 0.906 & 0.224 && 0.561 & 0.914 & 0.270 \\
			
			$Ground$ & 0.906 & 0.977 & 0.871 && 0.836 & 0.956 & 0.753 && 0.727 & 0.918 & 0.647 \\ \midrule
			
			$Speakers=150$\\
			
			$Train$ & 0.487 & 0.880 & 0.149 && 0.486 & 0.900 & 0.150 && 0.536 & 0.912 & 0.226 \\
			
			$Ground$ & 0.885 & 0.973 & 0.836 && 0.714 & 0.926 & 0.687 && 0.671 & 0.908 & 0.568 \\

			\bottomrule
		\end{tabular}
	\end{adjustbox}
\end{table}

The architecture requires a sufficient number of speech variations from users to explore the proper feature relationship between speech frames. Limiting the number of speakers to 25 caused the interpretation of speech to be reduced. Hence, the architecture struggles to find better speech relations, and lessened performance is observed. Increasing the number of speakers to 50 balances the speech variations in the training data and causes an increase in accuracy.

\section{Discussion}
\label{sec:future_work}

The pairwise architecture with training strategy results in a good performance in the speaker recognition process. However, throughout the investigation, the architecture tends to have some issues that have to be considered. Firstly, training the architecture with lesser speech variation causes overfitting, observed while keeping $speaker=25$. Secondly, as the augmentation procedure fuses noises, speech data with excessive noises may not generate a good result. \added{The degradation of performance due to excessive noise is a common concern, and by using a denoising mechanism, the situation can be handled \cite{azarang2020review}.} Further, as the system is fully segmentation dependent, the target lies in developing an optimal audio segmentation procedure. Resolving these challenges would benefit the architecture for a wide range of speaker recognition and evaluation usage.

\added{Apart from the limitations, the u-vector strategy requires no pre-training on large speaker datasets, which is often observed in i-vector, d-vector, and x-vector \cite{ohi2021deep}. Further, the u-vector strategy requires comparatively less per-speaker data than the other embedding strategies mentioned above. In the case of augmentations, u-vector architecture does not require any labeled data, which is primarily observed in generative speaker recognition architectures \cite{pal2021meta}. In an overall perspective, the u-vector mitigates the requirement of labeled data to a minimum.}

\added{Aside from implementing u-vectors in an unsupervised and semi-supervised manner, the strategy can be implemented in self-supervised learning. Although self-supervised learning has its various forms based on the domain, u-vector aligns with contrastive self-supervised learning strategies \cite{jaiswal2021survey}. In general, self-supervised learning learns better model representation (weights of a model) from unlabeled data. Further, the trained model is used to train on a small quantity of labeled data. U-vectors can be used to initialize model representations in the first stage of self-supervised learning. Further, a classification method/layer can be equipped with the model for training the model on labeled data.}

\added{Finally, industrial systems often can not label every data. However, if a large labeled dataset is required, then building a model can become costly. Therefore, unsupervised, semi/self-supervised learning in speaker recognition systems is expected to produce low-cost and attainable systems.}

\section{Conclusion}
\label{sec:conclusion}
The paper introduces a \added{system of generating} clusterable speech embedding based on the speakers, namely u-vector. The policy of the architecture deals with pseudo labels and trained from unlabeled datasets. The procedure is suitable for both semi\added{-}supervised and unsupervised training strategies. We evaluate such strategies with two appropriate deep learning architectures: pairwise and triplet. In the perspective of unlabeled data, the architecture performs at an acceptable rate concerning the number of speakers and speech segmentation errors. However, the method requires clean speech, and robust segmentation techniques to properly construct clusterable u-vectors, depending on speaker variations. We strongly believe that such an in-depth and hypothetical strategy of generating pseudo labels to train speaker recognition models would help researches develop new schemes.

\bibliographystyle{unsrt}
\bibliography{references.bib}

\begin{thebibliography}{10}

\bibitem{graves2006connectionist}
Alex Graves, Santiago Fern{\'a}ndez, Faustino Gomez, and J{\"u}rgen
  Schmidhuber.
\newblock Connectionist temporal classification: labelling unsegmented sequence
  data with recurrent neural networks.
\newblock In {\em Proceedings of the 23rd international conference on Machine
  learning}, pages 369--376, 2006.

\bibitem{azarang2020review}
Arian Azarang and Nasser Kehtarnavaz.
\newblock A review of multi-objective deep learning speech denoising methods.
\newblock {\em Speech Communication}, 122:1--10, 2020.

\bibitem{ning2019review}
Yishuang Ning, Sheng He, Zhiyong Wu, Chunxiao Xing, and Liang-Jie Zhang.
\newblock A review of deep learning based speech synthesis.
\newblock {\em Applied Sciences}, 9(19):4050, 2019.

\bibitem{fayek2017evaluating}
Haytham~M Fayek, Margaret Lech, and Lawrence Cavedon.
\newblock Evaluating deep learning architectures for speech emotion
  recognition.
\newblock {\em Neural Networks}, 92:60--68, 2017.

\bibitem{kamper2017embedded}
Herman Kamper, Karen Livescu, and Sharon Goldwater.
\newblock An embedded segmental k-means model for unsupervised segmentation and
  clustering of speech.
\newblock In {\em 2017 IEEE Automatic Speech Recognition and Understanding
  Workshop (ASRU)}, pages 719--726. IEEE, 2017.

\bibitem{o2018directions}
Mary~Grantham O’Brien, Tracey~M Derwing, Catia Cucchiarini, Debra~M Hardison,
  Hansj{\"o}rg Mixdorff, Ron~I Thomson, Helmer Strik, John~M Levis, Murray~J
  Munro, Jennifer~A Foote, et~al.
\newblock Directions for the future of technology in pronunciation research and
  teaching.
\newblock {\em Journal of Second Language Pronunciation}, 4(2):182--207, 2018.

\bibitem{ohi2021deep}
Abu~Quwsar Ohi, MF~Mridha, Md~Abdul Hamid, and Muhammad~Mostafa Monowar.
\newblock Deep speaker recognition: Process, progress, and challenges.
\newblock {\em IEEE Access}, 9:89619--89643, 2021.

\bibitem{kabir2021survey}
Muhammad~Mohsin Kabir, MF~Mridha, Jungpil Shin, Israt Jahan, and Abu~Quwsar
  Ohi.
\newblock A survey of speaker recognition: Fundamental theories, recognition
  methods and opportunities.
\newblock {\em IEEE Access}, 2021.

\bibitem{tiwari2010mfcc}
Vibha Tiwari.
\newblock Mfcc and its applications in speaker recognition.
\newblock {\em International journal on emerging technologies}, 1(1):19--22,
  2010.

\bibitem{chowdhury2019fusing}
Anurag Chowdhury and Arun Ross.
\newblock Fusing mfcc and lpc features using 1d triplet cnn for speaker
  recognition in severely degraded audio signals.
\newblock {\em IEEE Transactions on Information Forensics and Security},
  15:1616--1629, 2019.

\bibitem{ravanelli2018speaker}
Mirco Ravanelli and Yoshua Bengio.
\newblock Speaker recognition from raw waveform with sincnet.
\newblock In {\em 2018 IEEE Spoken Language Technology Workshop (SLT)}, pages
  1021--1028. IEEE, 2018.

\bibitem{9207519}
J.~A. {Chagas Nunes}, D.~{Macêdo}, and C.~{Zanchettin}.
\newblock Am-mobilenet1d: A portable model for speaker recognition.
\newblock In {\em 2020 International Joint Conference on Neural Networks
  (IJCNN)}, pages 1--8, 2020.

\bibitem{garcia2011analysis}
Daniel Garcia-Romero and Carol~Y Espy-Wilson.
\newblock Analysis of i-vector length normalization in speaker recognition
  systems.
\newblock In {\em Twelfth annual conference of the international speech
  communication association}, 2011.

\bibitem{variani2014deep}
Ehsan Variani, Xin Lei, Erik McDermott, Ignacio~Lopez Moreno, and Javier
  Gonzalez-Dominguez.
\newblock Deep neural networks for small footprint text-dependent speaker
  verification.
\newblock In {\em 2014 IEEE International Conference on Acoustics, Speech and
  Signal Processing (ICASSP)}, pages 4052--4056. IEEE, 2014.

\bibitem{snyder2018x}
David Snyder, Daniel Garcia-Romero, Gregory Sell, Daniel Povey, and Sanjeev
  Khudanpur.
\newblock X-vectors: Robust dnn embeddings for speaker recognition.
\newblock In {\em 2018 IEEE International Conference on Acoustics, Speech and
  Signal Processing (ICASSP)}, pages 5329--5333. IEEE, 2018.

\bibitem{bai2021speaker}
Zhongxin Bai and Xiao-Lei Zhang.
\newblock Speaker recognition based on deep learning: An overview.
\newblock {\em Neural Networks}, 2021.

\bibitem{wang2018unsupervised}
Qing Wang, Wei Rao, Sining Sun, Leib Xie, Eng~Siong Chng, and Haizhou Li.
\newblock Unsupervised domain adaptation via domain adversarial training for
  speaker recognition.
\newblock In {\em 2018 IEEE International Conference on Acoustics, Speech and
  Signal Processing (ICASSP)}, pages 4889--4893. IEEE, 2018.

\bibitem{garcia2014unsupervised}
Daniel Garcia-Romero, Alan McCree, Stephen Shum, Niko Brummer, and Carlos
  Vaquero.
\newblock Unsupervised domain adaptation for i-vector speaker recognition.
\newblock In {\em Proceedings of Odyssey: The Speaker and Language Recognition
  Workshop}, volume~8, 2014.

\bibitem{garcia2014improving}
Daniel Garcia-Romero, Xiaohui Zhang, Alan McCree, and Daniel Povey.
\newblock Improving speaker recognition performance in the domain adaptation
  challenge using deep neural networks.
\newblock In {\em 2014 IEEE Spoken Language Technology Workshop (SLT)}, pages
  378--383. IEEE, 2014.

\bibitem{schroff2015facenet}
Florian Schroff, Dmitry Kalenichenko, and James Philbin.
\newblock Facenet: A unified embedding for face recognition and clustering.
\newblock In {\em Proceedings of the IEEE conference on computer vision and
  pattern recognition}, pages 815--823, 2015.

\bibitem{ohi2020autoembedder}
Abu~Quwsar Ohi, MF~Mridha, Farisa~Benta Safir, Md~Abdul Hamid, and
  Muhammad~Mostafa Monowar.
\newblock Autoembedder: A semi-supervised dnn embedding system for clustering.
\newblock {\em Knowledge-Based Systems}, 204:106190, 2020.

\bibitem{zhang2018text}
Chunlei Zhang, Kazuhito Koishida, and John~HL Hansen.
\newblock Text-independent speaker verification based on triplet convolutional
  neural network embeddings.
\newblock {\em IEEE/ACM Transactions on Audio, Speech, and Language
  Processing}, 26(9):1633--1644, 2018.

\bibitem{campbell2006support}
William~M Campbell, Douglas~E Sturim, and Douglas~A Reynolds.
\newblock Support vector machines using gmm supervectors for speaker
  verification.
\newblock {\em IEEE signal processing letters}, 13(5):308--311, 2006.

\bibitem{kenny2007joint}
Patrick Kenny, Gilles Boulianne, Pierre Ouellet, and Pierre Dumouchel.
\newblock Joint factor analysis versus eigenchannels in speaker recognition.
\newblock {\em IEEE Transactions on Audio, Speech, and Language Processing},
  15(4):1435--1447, 2007.

\bibitem{dehak2010front}
Najim Dehak, Patrick~J Kenny, R{\'e}da Dehak, Pierre Dumouchel, and Pierre
  Ouellet.
\newblock Front-end factor analysis for speaker verification.
\newblock {\em IEEE Transactions on Audio, Speech, and Language Processing},
  19(4):788--798, 2010.

\bibitem{molla2004effectiveness}
KI~Molla and Keikichi Hirose.
\newblock On the effectiveness of mfccs and their statistical distribution
  properties in speaker identification.
\newblock In {\em 2004 IEEE Symposium on Virtual Environments, Human-Computer
  Interfaces and Measurement Systems, 2004.(VCIMS).}, pages 136--141. IEEE,
  2004.

\bibitem{ioffe2006probabilistic}
Sergey Ioffe.
\newblock Probabilistic linear discriminant analysis.
\newblock In {\em European Conference on Computer Vision}, pages 531--542.
  Springer, 2006.

\bibitem{chen2015locally}
Yu-hsin Chen, Ignacio Lopez-Moreno, Tara~N Sainath, Mirk{\'o} Visontai, Raziel
  Alvarez, and Carolina Parada.
\newblock Locally-connected and convolutional neural networks for small
  footprint speaker recognition.
\newblock In {\em Sixteenth Annual Conference of the International Speech
  Communication Association}, 2015.

\bibitem{snyder2017deep}
David Snyder, Daniel Garcia-Romero, Daniel Povey, and Sanjeev Khudanpur.
\newblock Deep neural network embeddings for text-independent speaker
  verification.
\newblock In {\em Interspeech}, pages 999--1003, 2017.

\bibitem{redko2019advances}
Ievgen Redko, Emilie Morvant, Amaury Habrard, Marc Sebban, and Youn{\`e}s
  Bennani.
\newblock {\em Advances in Domain Adaptation Theory}.
\newblock Elsevier, 2019.

\bibitem{spyrou2020data}
Evaggelos Spyrou, Eirini Mathe, Georgios Pikramenos, Konstantinos Kechagias,
  and Phivos Mylonas.
\newblock Data augmentation vs. domain adaptation—a case study in human
  activity recognition.
\newblock {\em Technologies}, 8(4):55, 2020.

\bibitem{guo2017learning}
Qing Guo, Wei Feng, Ce~Zhou, Rui Huang, Liang Wan, and Song Wang.
\newblock Learning dynamic siamese network for visual object tracking.
\newblock In {\em Proceedings of the IEEE international conference on computer
  vision}, pages 1763--1771, 2017.

\bibitem{ren2019semi}
Yazhou Ren, Kangrong Hu, Xinyi Dai, Lili Pan, Steven~CH Hoi, and Zenglin Xu.
\newblock Semi-supervised deep embedded clustering.
\newblock {\em Neurocomputing}, 325:121--130, 2019.

\bibitem{tan2020rvad}
Zheng-Hua Tan, Najim Dehak, et~al.
\newblock rvad: An unsupervised segment-based robust voice activity detection
  method.
\newblock {\em Computer Speech \& Language}, 59:1--21, 2020.

\bibitem{brent1999speech}
Michael~R Brent.
\newblock Speech segmentation and word discovery: A computational perspective.
\newblock {\em Trends in Cognitive Sciences}, 3(8):294--301, 1999.

\bibitem{yang2019deep}
Xu~Yang, Cheng Deng, Feng Zheng, Junchi Yan, and Wei Liu.
\newblock Deep spectral clustering using dual autoencoder network.
\newblock In {\em Proceedings of the IEEE Conference on Computer Vision and
  Pattern Recognition}, pages 4066--4075, 2019.

\bibitem{garofolo1993darpa}
John~S Garofolo, Lori~F Lamel, William~M Fisher, Jonathan~G Fiscus, and David~S
  Pallett.
\newblock Darpa timit acoustic-phonetic continous speech corpus cd-rom. nist
  speech disc 1-1.1.
\newblock {\em STIN}, 93:27403, 1993.

\bibitem{panayotov2015librispeech}
Vassil Panayotov, Guoguo Chen, Daniel Povey, and Sanjeev Khudanpur.
\newblock Librispeech: an asr corpus based on public domain audio books.
\newblock In {\em 2015 IEEE International Conference on Acoustics, Speech and
  Signal Processing (ICASSP)}, pages 5206--5210. IEEE, 2015.

\bibitem{kjartansson-etal-sltu2018}
Oddur Kjartansson, Supheakmungkol Sarin, Knot Pipatsrisawat, Martin Jansche,
  and Linne Ha.
\newblock {Crowd-Sourced Speech Corpora for Javanese, Sundanese, Sinhala,
  Nepali, and Bangladeshi Bengali}.
\newblock In {\em Proc. The 6th Intl. Workshop on Spoken Language Technologies
  for Under-Resourced Languages (SLTU)}, pages 52--55, Gurugram, India, August
  2018.

\bibitem{reddy2019scalable}
Chandan~KA Reddy, Ebrahim Beyrami, Jamie Pool, Ross Cutler, Sriram Srinivasan,
  and Johannes Gehrke.
\newblock A scalable noisy speech dataset and online subjective test framework.
\newblock {\em Proc. Interspeech 2019}, pages 1816--1820, 2019.

\bibitem{huang2017densely}
Gao Huang, Zhuang Liu, Laurens Van Der~Maaten, and Kilian~Q Weinberger.
\newblock Densely connected convolutional networks.
\newblock In {\em Proceedings of the IEEE conference on computer vision and
  pattern recognition}, pages 4700--4708, 2017.

\bibitem{hermans2017defense}
Alexander Hermans, Lucas Beyer, and Bastian Leibe.
\newblock In defense of the triplet loss for person re-identification.
\newblock {\em arXiv preprint arXiv:1703.07737}, 2017.

\bibitem{kingma2014adam}
Diederik~P Kingma and Jimmy Ba.
\newblock Adam: A method for stochastic optimization.
\newblock {\em arXiv preprint arXiv:1412.6980}, 2014.

\bibitem{pal2021meta}
Monisankha Pal, Manoj Kumar, Raghuveer Peri, Tae~Jin Park, So~Hyun Kim,
  Catherine Lord, Somer Bishop, and Shrikanth Narayanan.
\newblock Meta-learning with latent space clustering in generative adversarial
  network for speaker diarization.
\newblock {\em IEEE/ACM Transactions on Audio, Speech, and Language
  Processing}, 29:1204--1219, 2021.

\bibitem{jaiswal2021survey}
Ashish Jaiswal, Ashwin~Ramesh Babu, Mohammad~Zaki Zadeh, Debapriya Banerjee,
  and Fillia Makedon.
\newblock A survey on contrastive self-supervised learning.
\newblock {\em Technologies}, 9(1):2, 2021.

\end{thebibliography}
\end{document}